\documentclass{article}

\usepackage{arxiv}

\usepackage[utf8]{inputenc} 
\usepackage[T1]{fontenc}    
\usepackage{hyperref}       
\usepackage{url}            
\usepackage{booktabs}       
\usepackage{amsfonts}       
\usepackage{nicefrac}       
\usepackage{microtype}      
\usepackage{lipsum}		
\usepackage{graphicx}
\usepackage{natbib}
\usepackage{doi}
\usepackage{comment}

\title{Super-Resolved Image of M87 Observed with \linebreak East Asian VLBI Network}

\author{
\href{https://orcid.org/0000-0003-0236-0600}{\includegraphics[scale=0.06]{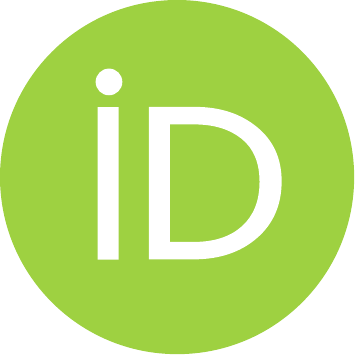}\hspace{1mm}Fumie Tazaki $^{1,}$},
\href{https://orcid.org/0000-0001-6311-4345}{\includegraphics[scale=0.06]{orcid.pdf}\hspace{1mm}Yuzhu Cui $^{2,3}$}, 
\href{https://orcid.org/0000-0001-6906-772X}{\includegraphics[scale=0.06]{orcid.pdf}\hspace{1mm}Kazuhiro Hada $^{4}$},
{\bf Motoki Kino $^{4,5}$}, \\
\And
\href{https://orcid.org/0000-0001-6083-7521}{\includegraphics[scale=0.06]{orcid.pdf}\hspace{1mm}Ilje Cho $^{6}$}, 
\href{https://orcid.org/0000-0002-4417-1659}{\includegraphics[scale=0.06]{orcid.pdf}\hspace{1mm}Guang-Yao Zhao $^{6}$}, 
\href{https://orcid.org/0000-0002-9475-4254}{\includegraphics[scale=0.06]{orcid.pdf}\hspace{1mm}Kazunori Akiyama $^{7,8,4}$}, 
Yosuke Mizuno$^{2,9,10}$,\\
\And
\href{https://orcid.org/0000-0002-7322-6436}{\includegraphics[scale=0.06]{orcid.pdf}\hspace{1mm}Hyunwook Ro $^{11,12}$}, 
Mareki Honma $^{4,13,14}$,
Ru-Sen Lu $^{15,16,17}$, 
Zhi-Qiang Shen $^{15,16}$,\\ 
\And
\href{https://orcid.org/0000-0003-0721-5509}{\includegraphics[scale=0.06]{orcid.pdf}\hspace{1mm}Lang Cui $^{18}$},
and Yoshinori Yonekura $^{19}$
\\
\\
$^{1}$ Tokyo Electron Technology Solutions Limited, Iwate 023-1101, Japan \\ {fumie.tazaki@tel.com}\\
$^{2}$ Tsung-Dao Lee Institute, Shanghai Jiao Tong University, Shanghai 201210, China\\
$^{3}$ Research Center for Intelligent Computing Platforms, Zhejiang Laboratory, Hangzhou 311100, China\\
$^{4}$ Mizusawa VLBI Observatory, National Astronomical Observatory of Japan, Iwate 023-0861, Japan\\
$^{5}$ Academic Support Center, Kogakuin University of Technology and Engineering, Tokyo 192-0015, Japan\\
$^{6}$ Instituto de Astrofísica de Andalucía-CSIC, Glorieta de la Astronomía s/n, E-18008 Granada, Spain\\
$^{7}$ Massachusetts Institute of Technology Haystack Observatory, Westford, MA 01886, USA\\
$^{8}$ Black Hole Initiative, Harvard University, Cambridge, MA 02138, USA\\
$^{9}$ School of Physics \& Astronomy, Shanghai Jiao Tong University, \\800 Dongchuan Road, Shanghai 200240, China\\
$^{10}$ Institut f\"ur Theoretische Physik, Goethe-Universit\"at Frankfurt,\\ Max-von-Laue-Stra{\ss}e 1, D-60438 Frankfurt am Main, Germany\\
$^{11}$ Department of Astronomy, Yonsei University, Seodaemun-gu, Seoul 03722, Republic of Korea\\
$^{12}$ Korea Astronomy and Space Science Institute, Yuseong-gu, Daejeon 34055, Republic of Korea\\
$^{13}$ Department of Astronomical Science, The Graduate University for Advanced Studies, SOKENDAI,\\Tokyo 181-8588, Japan\\
$^{14}$ Institute of Astronomy, The University of Tokyo, Tokyo 181-0015, Japan\\
$^{15}$ Shanghai Astronomical Observatory, Chinese Academy of Sciences, Shanghai 200030, China\\
$^{16}$ Key Laboratory of Radio Astronomy, Chinese Academy of Sciences, Nanjing 210008, China\\
$^{17}$ Max-Planck-Institut für Radioastronomie, Auf dem Hügel 69, D-53121 Bonn, Germany\\
$^{18}$ Xinjiang Astronomical Observatory, CAS, 150 Science 1-Street, Urumqi 830011, China\\
$^{19}$ Center for Astronomy, Ibaraki University, Ibaraki 310-8512, Japan\\
}

\date{}

\renewcommand{\headeright}{}
\renewcommand{\undertitle}{}
\renewcommand{\shorttitle}{Super-resolution Image of M87 EAVN}

\begin{document}
\maketitle
\begin{abstract}
    Obtaining high-resolution images at centimeter-or-longer wavelengths is vital for understanding the physics of jets. We reconstructed images from the M87 22\,GHz data observed with the East Asian VLBI Network (EAVN) by using the regularized maximum likelihood (RML) method, which is different from the conventional imaging method CLEAN. Consequently, a bright core and jet extending about 30\,mas to the northwest were detected with a higher resolution than in the CLEAN image. The width of the jet was 0.5\,mas at 0.3\,mas from the core, consistent with the width measured in the 86\,GHz image in the previous study. In addition, three ridges were able to be detected at around 8\,mas from the core, even though the peak-to-peak separation was only 1.0\,mas. This indicates that the RML image's spatial resolution is at least 30\% higher than that of the CLEAN image. This study is an important step for future multi-frequency and high-cadence observations of the EAVN to discuss the more detailed structure of the jet and its time variability.
\end{abstract}

\keywords{active galactic nuclei; jet; very-long-baseline interferometry; M87; imaging}

\section{Introduction}

\textls[-25]{M87, located at 16.8 Mpc from Earth in the constellation Virgo, is a giant elliptical galaxy with a super-massive black hole of $6.5 \times 10^9 \,M_\odot$ at its center \citep{eht2019}.
This close proximity to a huge black hole allows for a linear resolution of 1 milliarcsecond (mas) = 0.08 pc = 130 Schwarzschild radii ($R_{\rm s}$).}
The relativistic jet erupts from the central core, emitting radiation ranging from radio to X-rays and gamma rays \citep{mwl2021}.
Even considering only the radio band, the jet's appearance is quite different depending on the wavelength.
For VLBI at centimeter-or-longer wavelengths, M87 jets extending to $\sim$900\,mas at 150\,mm \citep{Hada2012} and $\sim$20\,mas at 13\,mm~\citep{Hada2017} have been observed; however, as the wavelength becomes shorter, the downstream of the jet becomes less visible, extending to $\sim$10\,mas at 7\,mm \citep{Walker2018} and $\sim$3\,mas at 3.5\,mm \citep{Had16, Kim2018}.
The 1.3\,mm data observed with the EHT imaged the ring-like structure in the very vicinity of the black hole \citep{eht2019}.
The extended jet at the 1.3 mm waveband is so faint that it was not detected in the 2017 EHT observation. However, by improving short-baseline coverage, the dynamic range of the image can be increased, and the faint, extended jet can be recovered.
How far downstream the jet can be detected depends on the sensitivity of the telescope, but there is no doubt that observations at centimeter-or-longer wavelengths are still more suitable for studying the downstream of the jet.

It is more difficult to obtain high-resolution images downstream of a jet than upstream because shorter wavelengths have a higher spatial resolution for the same telescope aperture.
One method for obtaining high-resolution images of the downstream of a jet is space VLBI. By connecting ground and space telescopes, VLBI can be constructed to obtain a long baseline length.
This effectively means that it plays the same role as a giant telescope that is larger than the size of the Earth.
For example, the VLBI Space Observatory Programme (VSOP) satellite, led by the Institute of Space and Astronautical Science (ISAS) in JAXA in collaboration with the National Astronomical Observatory of Japan (NAOJ), was launched in 1997 and renamed HALCA, which carries a radio telescope with an 8-meter diameter~\citep{Hirabayashi2000}. High spatial resolution observations at 1.6 GHz and 5 GHz using VSOP successfully resolved the jet into three ridges at the mas scale \citep{Asa16}. Investigating how the complex internal structure of the jet evolves over time is critical to understanding the jet's physics.
However, as space VLBI observation time is limited, it is difficult to conduct studies that closely monitor the jet and reveal component motions in detail.

Therefore, to image the jet's fine structure further from the central region, it is necessary to observe the jet at centimeter-or-longer wavelengths, where the jet is the brightest, and to reconstruct a high-resolution image.
The regularized maximum likelihood (RML) method applied in this study is an imaging technique for interferometric data that was primarily developed for imaging the data obtained with the Event Horizon Telescope (EHT).
While CLEAN, a conventional method, creates an image model after constructing a dirty map, resulting in images with resolutions limited by the beam size, RML directly constructs an image model that fits the observed data,  resulting in higher-resolution images than the nominal diffraction limit determined by the \emph{uv} coverage.
As the ideal spatial resolution of VLBI is several times smaller than beam size \citep{Honma2014}, the use of RML may yield images with several-times-higher spatial resolutions than CLEAN images.

The East Asian VLBI Network (EAVN) is a VLBI system in East Asia, and currently consists of up to 16 telescopes with a longest baseline of 5078 km \citep{Waj22}.
Three science working groups (SWGs), active galactic nuclei (AGNs), evolved stars, and star-forming regions are operated in the EAVN, of which M87 has continued to be monitored since 2017 at 13 and 7~mm wavelengths \citep{Cui21} as a target of the AGN SWG.
This is a major monitoring project that has been carried out since the era of the KVN and VERA Array (KaVA; VLBI network in Japan and Korea; \citep{Niinuma2014}), and the results of KaVA's M87 observations are summarized in \citet{Park2019}, which found where the jet accelerated and transitioned from subluminal to superluminal speeds.
These high-cadence monitoring observations provide precise measurements of jet motion, which is an important step toward understanding the physics and evolution of jets.

\section{Observations and Data Reductions}

The data treated in this paper were observed with the EAVN on 18 March 2017, and have already been published in \citet{Cui21}.
The paper \citep{Cui21} presents observations of four AGNs in two frequency bands of the EAVN, namely, 22 GHz and 43 GHz.
They used the Astronomical Image Processing System \citep[AIPS;][]{Greisen2003} provided by the National Radio Astronomy Observatory (NRAO) for an initial calibration of the complex visibility. After that, CLEAN imaging and self-calibration were performed from the calibrated data using DIFMAP software \citep{Shepherd1994}.
On the other hand, the data treated in this paper are only the M87 22 GHz observation, which is imaged with a different imaging method (see Section \ref{sec:3} for the details) using the initial calibrated data prior to imaging.
In this section, a brief overview of the observational data used in this study is given.

\begin{figure}[h]
\includegraphics[width=10.5 cm]{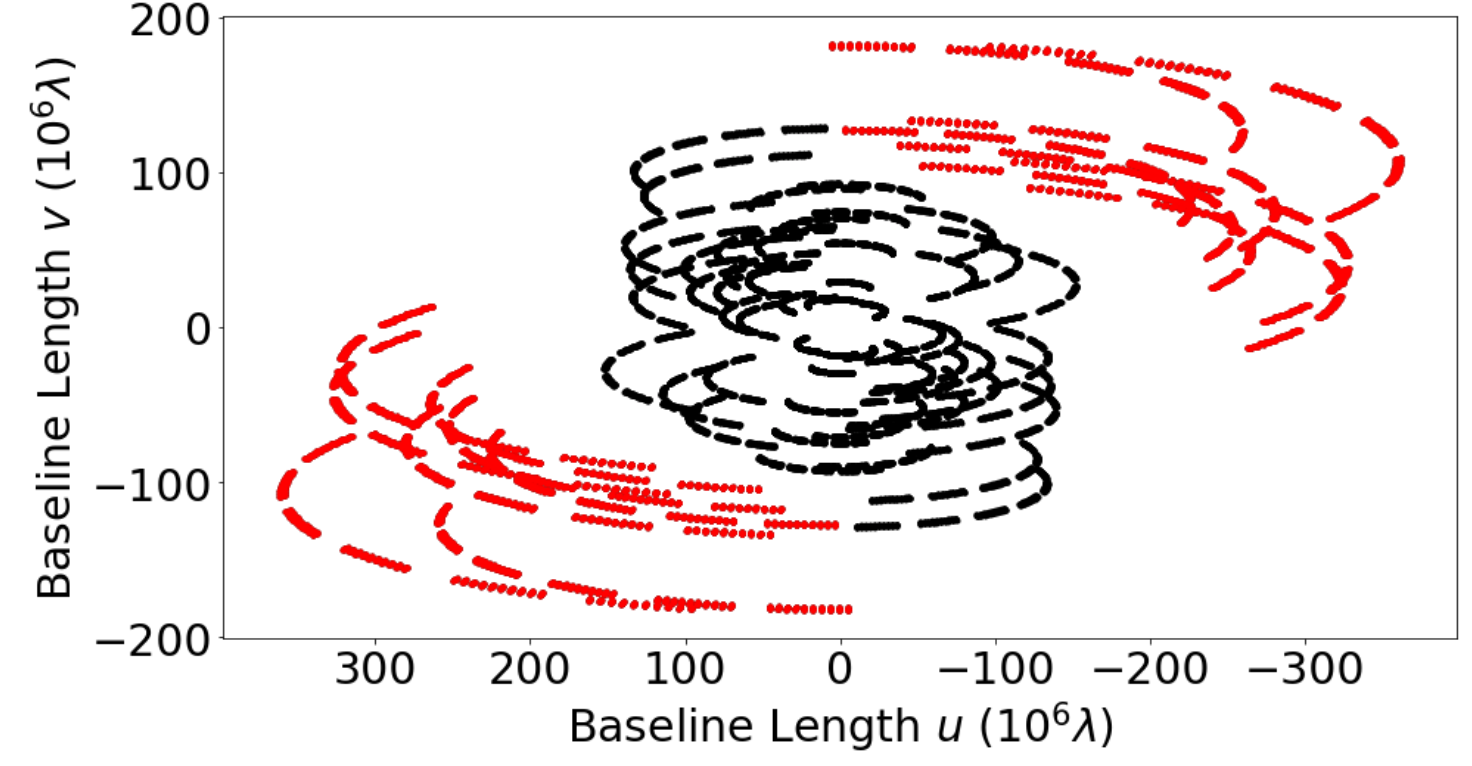}
\caption{The \emph{uv}
coverage for the M87 session. Red curves indicate baselines related to NSRT.\label{uvcover}}
\end{figure}

The telescopes participating in this observation were the KVN and VERA Array (KaVA), Tianma 65 m Radio Telescope (TMRT), Nanshan 26 m Radio Telescope (NSRT), and Hitachi (HIT).
VERA consists of four stations in Japan, namely, Mizusawa (MIZ), Iriki (IRK), Ogasawara (OGA), and Ishigakijima (ISG). KVN stands for Korean VLBI Network, which has three stations, namely, Yonsei (KYS), Ulsan (KUS), and Tamna (KTN), though KYS was unable to participate in this observation due to an issue at the site. See Table 1 in~\citet{Cui21} for specifications of each participating telescope.
Figure~\ref{uvcover} shows that NSRT contributes to filling the outer part of the \emph{uv} plane.
The longest baseline is 5100 km between NSRT and Ogasawara, which corresponds to the smallest synthetic beam size of 0.6\,mas in the northwest--southeast direction.

M87 was observed along with 3C~273, 1219+044, and M84. Within the 7-hour session, M87 was observed for 6 scans at 47 min per scan.
The recording rate was 1\,Gbps (2-bit sampling), where a total bandwidth of 256 MHz was divided into eight 32 MHz intermediate-frequency (IF) bands.
Only left-hand circular polarization was received.
All the data were correlated at the Daejeon hardware correlator installed at the Korea Astronomy and Space Science Institute (KASI).
The initial calibration of visibility amplitude, bandpass, and phase was performed in the standard manner by using the AIPS software package.

\section{Imaging}\label{sec:3}

We reconstructed M87 images from the EAVN data set with the regularized maximum likelihood (RML) method implemented in the Sparse Modeling Imaging Library for Interferometry \citep[SMILI;][]{Akiyama2017a,Akiyama2017b}, which is developed primarily for imaging EHT data \citep{eht2019IV}.
The VLBI observation fills the \emph{uv}-plane sampling depending on the telescope placement and the position of the source, as shown in Figure~\ref{uvcover}.
To obtain an image from these data, the conventional method CLEAN draws a dirty map by inverse Fourier transform with zeros in the empty \emph{uv} plane and reconstructs a set of point-source models from the dirty map.
RML, on the other hand, uses regularization to ensure a plausible image consistent with the data without performing an inverse Fourier transform.
In this imaging, we use the regularization terms of the weighted-L1 (wL1), total variation (TV), total squared variation (TSV), and maximum entropy method (MEM) \citep[see Appendix A in][for the definition of each term]{eht2019IV},
which describe assumptions about the source structure, such as that only limited regions in the field of view have brightness or that the source structure is smooth.
Each term contains a hyperparameter, which adjusts the relative weighting of the regularization term to the data.
If the weight of the regularization term is too strong, the image will be inconsistent with the data.
On the other hand, if the weight of the data is too strong, the image will not reflect the features of the image expressed by the regularization term.
This allows the production of dirty beam-free images with RML, which can recover finer structures than CLEAN.

\begin{figure}[h]
\includegraphics[width=14.5 cm]{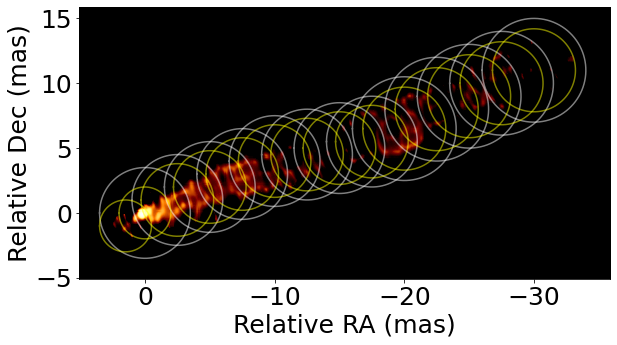}
\caption{The averaged image of 48 final images (see Figure~\ref{all_image}) and regions of the window setting. White circles represent areas where pixel values are restored during image reconstruction. Flux densities generated outside the yellow circles by the image reconstruction are considered noise, and pixel values are set to zero.\label{window}}
\end{figure}

We used visibility amplitudes, log closure amplitudes, and closure phases for image reconstruction.
The visibility amplitude of TMRT shows some offsets compared with that of KaVA.
Due to the uncertainty of a priori calibration, the visibility amplitudes of NSRT and HIT are very low compared with KaVA.
These systematic errors in visibility amplitude do not affect the log closure amplitude because they are offset in the process of calculating the log closure amplitude.
Therefore, the baselines, including TMRT, NSRT, and HIT, are excluded only from the visibility amplitude for the first imaging, and they are applied after the visibility amplitude is corrected by self-calibration.

wL1 regularization applies a pixel intensity penalty so that the dark areas in a prior image are also dark in the restored image, reducing the noise of the background region.
The prior image is obtained by convolving the CLEAN map with 1.5 mas circular Gaussian.
Since only loose constraints on the intensity distribution of the core and jet are required, we convolved the map with a large-enough circular Gaussian; therefore, our images are not affected by the detailed CLEAN model.
We applied a total flux of 2.0\,Jy based on the CLEAN results.
The image properties are set to a pixel size of 80 $\mu$as and a field of view of 41\,mas $\times$ 20\,mas, referring to Figure 9 of \citet{Cui21}.
In addition, the image window, in which the intensity is calculated in imaging, was set as shown by the white circles in Figure~\ref{window}.
To remove noise within the image window and outside of the source structure, the brightness outside of the yellow-circles region in Figure~\ref{window} was corrected to zero at the end of each image.
As the jet's structure is restored in an area much smaller than these windows, it can be assumed that the window settings are appropriate and that the subsequent image reconstruction is successful.

\begin{figure}[h]
\includegraphics[width=16 cm]{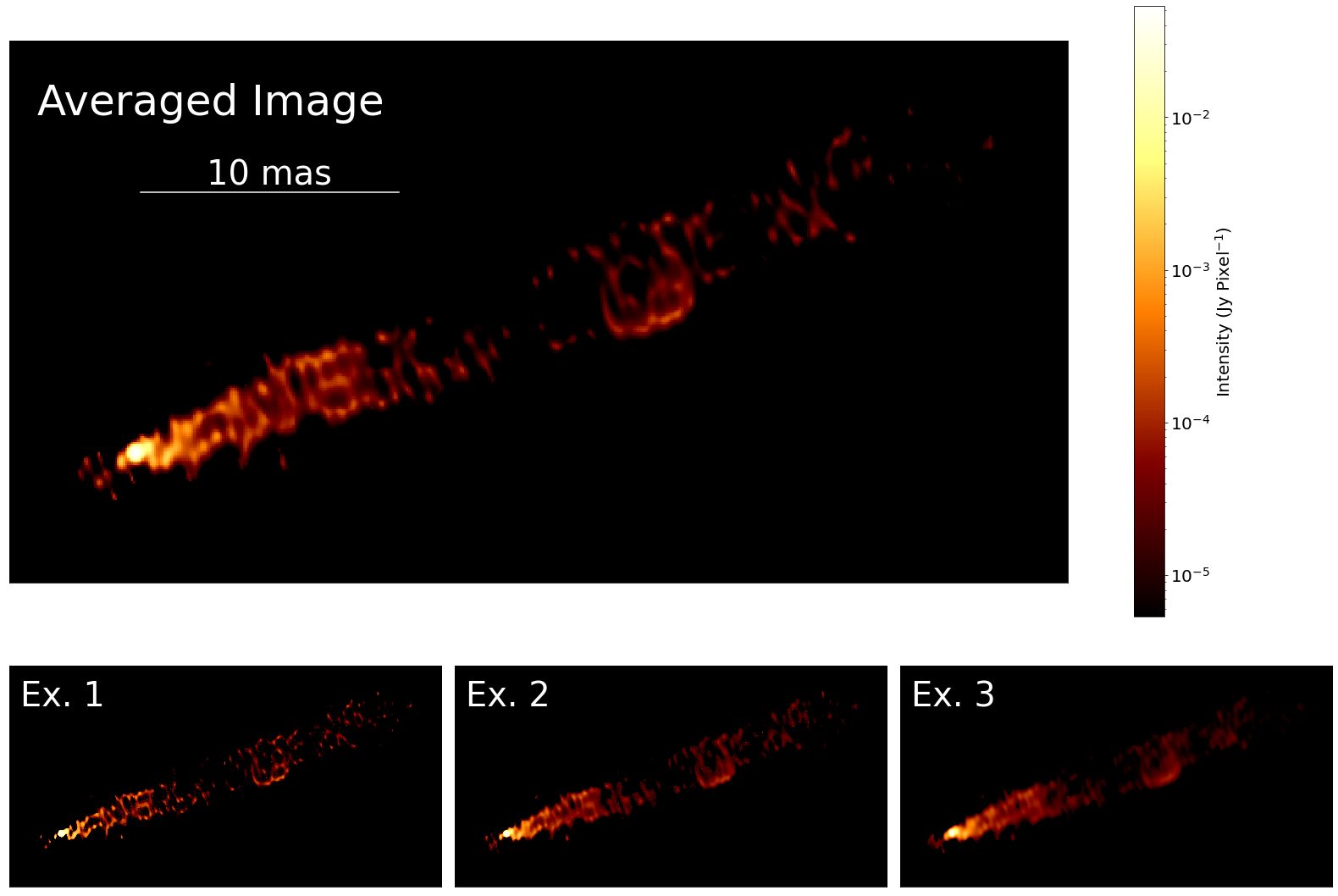}
\caption{The top panel shows the averaged image of the final 48 images. The bottom panel shows \linebreak 3 images taken from a total of 48 images as examples to see how much difference there is between the selected 48 images. \label{all_image}}
\end{figure}

In the first step, after excluding the HIT, NSRT, and TMRT baselines, we added a 200\% error to the visibility amplitudes to reduce the weight of the visibility amplitudes and reconstruct images with a greater emphasis on the closure quantities information.
We started the first imaging with the initial image of a circular Gaussian with the full width at a half maximum of 0.1\,mas.
The regularization parameters of wL1, TV, TSV, and MEM were set to 1, 10, 10, and 0.0001, respectively.
We performed the first self-calibration by using the obtained image as a model and obtained a calibrated \emph{uv} data set to proceed to the next step of the parameter survey.
Self-calibration restored the depressed visibility amplitudes, and all baselines were used hereafter.

In the next step, iterative imaging was performed using the self-calibrated \emph{uv} data.
The iterative pipeline of imaging and self-calibration was created to investigate how the image changes or is consistent with the data depending on the parameter set.
The following parameter combinations were prepared: additional errors of visibility amplitudes \linebreak (err) = [0.1,0.01,0], regularization parameters of MEM ($\lambda_{\rm MEM}$) = [0.01,0.001,0.0001],
regularization parameters of wL1 ($\lambda_{\rm wL1}$) = [10,1,0.1], and regularization parameters of TSV ($\lambda_{\rm TSV}$) = [100,10,1].
After the iterative imaging of 81 combinations of the parameter sets, we selected the final images with good fits to the data, i.e., those with reduced $\chi^2$ around 1.
The selection criteria of reduced $\chi^2$ for the closure phases and log closure amplitudes are less than 1.2 and 1.3, respectively.

\section{Image Properties}

In total, 48 final images, which fit the data better than the selection criteria, were selected among the 81 images.
There are slight differences among the 48 selected images depending on the parameter set. For example, MEM had a particularly pronounced effect on the images, with the larger $\lambda_{\rm MEM}$ resulting in a blurrier image. Ex. 1, 2, and 3 in Figure~\ref{all_image} show images with $\lambda_{\rm MEM}$ set to 0.0001, 0.001, and 0.01, respectively, all with a reduced $\chi^2$ near to 1. The top image in Figure \ref{all_image} is the average of 48 such images that are slightly different but meet the reduced $\chi^2$ selection criteria. All images have a bright central core from which a fainter jet extends in a northwesterly direction.
Furthermore, the jet dims at about 15\,mas from the core and brightens again at about 20\,mas, which is consistent with the CLEAN image \citep{Cui21}.

A counter-jet-like structure is also visible, though it is very faint and short compared with the approaching jet. We tried to calculate the noise level of the image at a sufficient distance from the source structure using DIFMAP, assuming that the pixel values of the averaged RML image are the CLEAN model. Using natural weighting, we estimated\linebreak  rms = 0.3 mJy/beam with the beam size (FWHM) of $1.39 \times 0.605$ mas in 11.6 degrees. The counter-jet-like structure was about six times brighter than the noise rms and is considered to be significantly detected. The brightest part of the counter feature in the average RML image, which is located 1.7 mas southeast of the core, has about 23\% of the brightness of the main jet symmetrically located relative to the core. This means that the brightness ratio (BR) of the main jet to the counter jet is about 4.3. However, there is no conspicuous blob to the northwest of the core that fully corresponds to the blob to the southeast.
In previous studies, the BR of the M87 jet at 1 mas from the core was about 5--25 \citep{Had16}, and at 0.5--3.1 mas, 10--15 \citep{Kov07}. In \citet{Ly07}, the BR integrated over a region was 14.4, but the peak-to-peak ratio was 3.4. Although this study is roughly consistent with the results of previous studies, more precise BR measurements are needed to estimate the jet velocity.

\begin{figure}[h]
\includegraphics[width=16 cm]{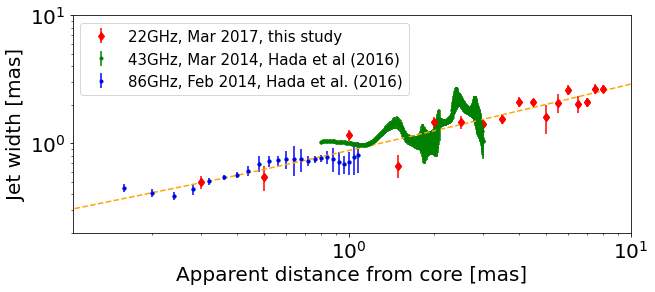}
\caption{The width of the jet relative to the apparent distance from the core is shown on a double-logarithmic graph. Red diamonds show the result of this study. Green and blue circles are obtained from 43 and 86\,GHz images, respectively, observed with VLBA and GBT in 2014 \citep{Had16}. The best-fit model ($W \propto r^{0.54 \pm 0.09}$) for the relationship between the apparent distance $r$ and jet width $W$, obtained from the images of this study, is shown with an orange dashed line. The observation epochs are noted in the legend.\label{width}}
\end{figure}

To evaluate the jet shape, all 48 images were processed as follows.
We rotated all images 19 degrees clockwise so that the jet extended horizontally rightward from the bright core. The vertical profile of this image was obtained to analyze the intensity distribution in the direction perpendicular to the jet axis. The profiles were located 0.3\,mas and 0.5\,mas from the core, and then in 0.5 mas increments up to 8.0 mas, they were averaged by $\pm$0.2\,mas horizontally, respectively.
The jet width was defined using the full width at half-maximum intensity (FWHM). If there are multiple peaks in the profile, the FWHM of the peaks at both ends and the peak-to-peak distance between them were used to calculate the width of the entire jet.
Jet widths were measured in all 48 images, and the average values are plotted in Figure~\ref{width} with red diamond marks. The error bars are their standard deviations. The widths of 0.3\,mas and 8\,mas from the core were 0.5\,mas and 2.7\,mas, respectively.
The jet width's dependence on the core distance can be fitted with a power law with an exponent of $0.54 \pm 0.09$ (95\% confidence level).
These widths are consistent with the jet width of the 43 and 86\,GHz images obtained with the CLEAN method in \citet{Had16}, and they are shown with green and blue circle marks in Figure~\ref{width}.

\begin{figure}[h]
\includegraphics[width=16 cm]{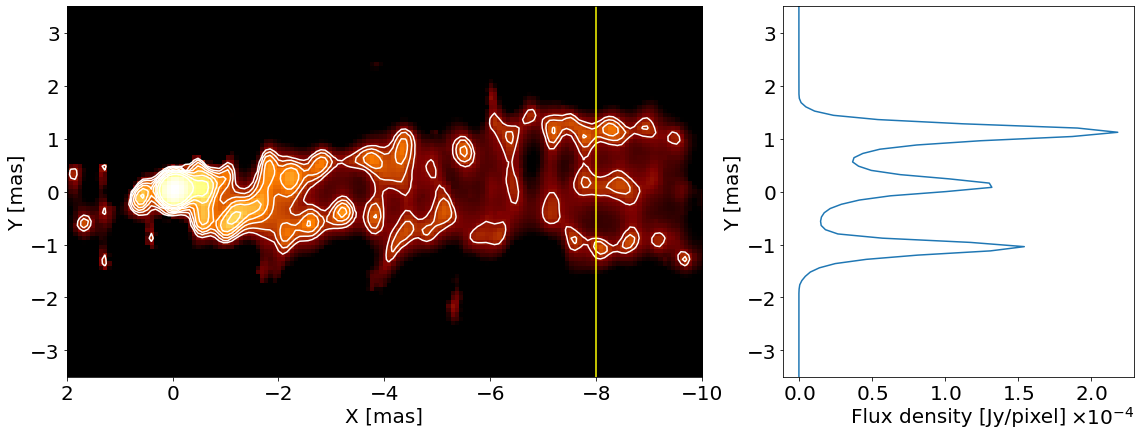}
\caption{\textbf{Left}: the average image rotated clockwise by 19 degrees. The lowest level of the contour line is 0.13\% of peak flux. The yellow vertical line is drawn at 8.0\,mas from the core, where the slice profile is investigated. \textbf{Right}: Slice profile integrating a region of $8.0 \pm 0.2$\,mas from the core.\label{slice_prof}}
\end{figure}

To see if we can resolve the ridge structure in the images, we investigated the profile perpendicular to the jet axis at a distance of 8\,mas along the jet from the core (Figure~\ref{slice_prof}), where we see three peaks in each image.
The best-fit parameters were obtained by fitting the profile with three Gaussian components.
The separation between the central and southern peaks was 1.1--1.4\,mas, and the separation between the northern and central peaks was even narrower at 0.9--1.1\,mas.
On the other hand, no triple ridges were detected in the CLEAN image obtained using the same observation data \citep{Cui21}. As the beam size used for the convolution of the CLEAN image was 1.35\,mas perpendicular to the jet, by assuming it to be the spatial resolution of the CLEAN image in this direction, the RML image with a peak separation of 1.0\,mas shows at least 30\% higher spatial resolution than the CLEAN image.
The ridges with the brightest peak intensities were, in order, the northern, the southern, and the central ridges.
The central ridge seems to be systematically thicker than the northern and southern ridges; however, for some images, the difference in width is so small that it is unclear whether there is a significant difference.
Previous observations at 1.6 GHz with VSOP \citep{Asa16} and the ultra-deep image of 2 Gbps VLBA with the phased VLA at 15 GHz \citep{Had17b} detected a triple-ridge structure more than 5 mas downstream from the core.
The triple ridge at 8 mas from the core detected in this study is also consistent in position and width with these, so it is likely that the same phenomenon was captured.

\section{Summary}

We reconstructed M87 images from the EAVN 22 GHz data set with the RML method implemented in the SMILI tool using sparse modeling techniques. Visibility amplitudes, log closure amplitudes, and closure phases are utilized for the image reconstruction. By setting various parameters in an iterative pipeline of imaging and self-calibration, 81 images were obtained. Finally, 48 images with a reduced $\chi^2$ of about 1 were selected as the final images.
Although there are slight differences among the selected images, depending on the parameter set, all images have a bright central core from which a fainter jet extends in a northwesterly direction. A counter-jet-like structure is also visible with the detection at 6$\sigma$, although it is very faint and short compared with the approaching jet.  We successfully measured the jet width of the region of a few mas at the root of the jet, which is consistent with the results of previous studies measured at 43 GHz and 86 GHz. We observed three peaks in the profile perpendicular to the jet axis at a distance of 8 mas from the core in each image, which is consistent with the previous studies from a space VLBI by VSOP and an ultra-deep image by the 2 Gbps VLBA observation with the phased VLA.
The image of the M87 jet obtained by RML allows us to examine the differences in the brightness and thickness of the three ridges in detail, as well as their distance from each other, features which could be lost by convolving with nominal beam sizes, as in the CLEAN method.

In the future, we must image the M87 jet at a centimeter-or-longer wavelength with higher resolution and sensitivity to investigate the entire jet from the root to the downstream with more precise profiles. This will help us to understand the acceleration and collimation mechanisms of the jet. If we observe the same region by using multiple wavelengths, we should be able to determine the physical properties, such as the optical depth and magnetic field. Furthermore, if monitoring observations are made at a resolution high enough to distinguish the three ridges, it is possible to study the time variation of the peculiar structure. Expanding the VLBI array will be one of the keys to achieving these goals.
An attempt to create a global array centered on East Asia has already begun with EATING VLBI, a joint effort between the EAVN and the Italian VLBI Network \citep{Hada2020}.
Collaboration between East Asia and Australia has also started as a joint observation between the Tidbinbilla 70 m radio telescope and the EAVN, and it will include other Australian stations in the future.
In addition, a radio telescope in Thailand is scheduled to join the EAVN in the future.
These array extensions will play a pilot role for the next generation of VLBI, such as SKA-VLBI~\citep{Dewdney2009} and ngVLA \citep{Murphy2018}.

\textbf{Funding:}
This research is funded by the following: JSPS (Japan Society for the Promotion of Science) Grant in Aid for Scientific Research (KAKENHI) (A) 22H00157 and (B) 18KK0090 (K.H.). K.H. is also funded by the Mitsubishi Foundation (201911019). Y.C. is funded by the China Postdoctoral Science Foundation (2022M712084). K.A. is financially supported by grants from the National Science Foundation (AST-1935980, AST-2034306, AST-2107681, AST-2132700, OMA-2029670). Y.M. is supported by the National Natural Science Foundation of China (12273022) and the Shanghai pilot program of international scientists for basic research (22JC1410600). R.-S.L. is supported by the Max Planck Partner Group of the MPG and the CAS, the Key Program of the NSFC (No. 11933007), the Key Research Program of Frontier Sciences, CAS (No. ZDBS-LY-SLH011), and the Shanghai Pilot Program for Basic Research---CAS, Shanghai Branch (No. JCYJ-SHFY-2022-013). L.C. is supported by the CAS ``Light of West China'' Program (No. 2021-XBQNXZ-005) and the NSFC (No. U2031212 and~61931002).

\textbf{Acknowledgments:}
This work made use of the East Asian VLBI Network (EAVN), which is operated under cooperative agreement by the National Astronomical Observatory of Japan (NAOJ), Korea Astronomy and Space Science Institute (KASI), Shanghai Astronomical Observatory (SHAO), and Xinjiang Astronomical Observatory (XAO). The operation of the Hitachi 32 m telescope is partially supported by the inter-university collaborative project ``Japanese VLBI Network (JVN)'' of NAOJ. We acknowledge all staff members and students who supported the operation of the array and the correlation of the data.

\end{document}